# Reinterpreting economic complexity in multiple dimensions

27 August 2024

Önder Nomaler* & Bart Verspagen**

* UNU-MERIT, Maastricht, the Netherlands, nomaler@merit.unu.edu

** Maastricht University, School of Business and Economics, the Netherlands, b.verspagen@maastrichtuniversity.nl

**Abstract**

We build on the interpretation of the Economic Complexity method as Correspondence Analysis (CA), and propose that the Canonical form of CA (CCA), which originated in the ecology literature, can be used to calculate multi-dimensional economic complexity. The traditional (CA) way of calculating economic complexity includes no "external" information such as countries' development characteristics to facilitate interpretation of "complexity". This has led to a wide range of fairly ad hoc interpretations of economic complexity on the basis of ex-post correlation to a long list of other variables. By the ex-ante inclusion of a number of country variables in the construction of the complexity indicators, CCA enables better interpretation, also in the case of multi-dimensional indicators. The analysis is further facilitated by another element of the ecologists' toolbox, the so-called biplots, which are CCA-based graph embeddings that represent a lower-dimensional product-space in which products and countries are positioned together, in mutual correspondence to each other. We show that in this way, CCA provides a richer account of development in many of its aspects, especially economic growth.

**INTRODUCTION**

The notion of economic complexity has been coined by César Hidalgo and Ricardo Hausmann (1) to capture the relationship between trade specialization of countries and their development level. Following the introduction of a public database and set of visualization tools (2), indicators based on this idea have been used widely in economic research. The term "complexity" refers to traded products, such as wheat, iron ore, cars or textiles machines, and the "economic complexity index" is a measure of the average complexity of the products that a country exports with above-average intensity. The way in which these indicators are defined has little to do with the common meaning of the term "complexity", which, by fairly basic intuitive reasoning, is adopted to reflect the level of productive capabilities that is supposedly needed to export certain products successfully.

A large body of literature has correlated economic complexity with a range of other indicators, suggesting that economic complexity is a central indicator related to many different dimensions of development: economic growth and income equality (3), greenhouse gas (GHG) emissions (4), health outcomes (5), unemployment rates (6), etc. Economic growth seems to be of particular interest as a variable that correlates to complexity (e.g., 7-9 to mention only a few; see also 10 for an overview). The wide variety of phenomena that are correlated to economic complexity does not help in obtaining a precise interpretation of the complexity indicators, but rather suggests that it is a fairly general indicator of which the exact working is hard to pin down.

This seems understandable from the interpretation of the economic complexity method that is offered by (11) and (12). They notice that the way in which economic complexity is calculated is identical to correspondence analysis (CA), which is an ordination method that is popular in various scientific fields, especially quantitative ecology. CA is an exploratory data analysis method, similar to principal component analysis or redundancy analysis (13), used to order quantities of different nature into a low-dimensional latent space, with the aim to describe the main tendencies for correlation in the data. An important element of CA is dimensionality reduction, e.g., in the case of economic complexity, from several thousands of traded products to the single dimension of the economic complexity index.

The salient difference of CA from other statistical methods for dimensionality reduction is that it is designated to be used on categorical (rather than continuous) data organized in the form of a two-way table. Ecologists often use CA to describe the (co-)occurrence patterns of species in particular locations (see 13, for both methods and examples). For example, CA could summarize a large database on the occurrence of different kinds of birds in many different sites, yielding a 2-dimensional space where bird types that tend to occur in the same locations are plotted close together. Such a "biplot" would contain both the different bird species, and the sites at which the species occur. Sites that are more similar in terms of the species they host appear closer to each other (as well as to the species they host) on the plot. In terms of how economic complexity uses CA, the role of species is played by products, and the role of sites by countries (or regions).

In canonical correspondence analysis (CCA), ecologists extend CA by the incorporation of an additional set of variables which capture physical characteristics of the locations, e.g., average temperature, altitude or humidity. These so-called environment variables are included into the analysis ex-ante, i.e., the summary dimensions are derived taking into account the physical characteristics of sites, rather than ex-post, e.g., when we would try to correlate the results of CA to the physical characteristics variables.

This suggests a direct linkage to the economic complexity literature: rather than correlating the economic complexity index derived by CA to a range of variables at the country level ex-post, we can include the specific variables of interest ex-ante, perform CCA, and derive an economic complexity indicator that is tailored to our specific research interests. This is what we will illustrate in this paper, by performing CCA on the trade data that the economic complexity field also uses, and by letting some of the many variables that economic complexity has been correlated with take the role of the environment variables.

We further propose to follow the example of quantitative ecology and use more than just a single indicator for economic complexity. CA reduces the highly-dimensional product space to an ordered range of summary dimensions, where each next summary dimension "explains" ever-fewer of the variance in the original trade data (this is explained in the next section below). CCA also offers multiple summary dimensions, but here the number of those is limited to the number of environmental variables that have been included in the analysis. There is no obvious reason why one should limit the analysis to just one summary indicator (as the economic complexity field commonly does), other than the fact that multiple dimensions (especially >2) will be harder to visualize and harder to interpret. If one dimension of economic complexity is hard to interpret, two or three will be even harder. However, CCA potentially addresses the interpretation problem by the inclusion of the environmental variables that reflect the researcher's ex-ante hypotheses on the factors that are likely to yield an interpretable narrative.

An alternative measure for complexity has been proposed in (14). (15) applies this measure to predict economic growth, i.e., the purpose of the alternative measure is similar to the interpretation of the Hidalgo and Hausmann complexity measure. While Hidalgo and Hausmann's

CA measure is ultimately based on eigenanalysis, the alternative measure (which results from a non-linear iterative algorithm) is not. Accordingly, the alternative measure is limited to reduce dimensionality to one. Thus, this alternative in (14) is beyond the scope of our paper which aims to emphasize the added value in looking beyond a single dimension in a strand of literature that is primarily aimed at dimensionality reduction.

**ECONOMIC COMPLEXITY AND CORRESPONDANCE ANALYSIS**

Economic Complexity Analysis starts with export value data in US\$ for $n$ products exported by $m$ countries. The revealed comparative advantage (RCA) of country $p$ in product $q$ is calculated as $R_{qp} = {s_{qp}}/{s_{+p}}$, where $s_{qp}$ is the share of country $p$ in total export value (over all countries) of product $q$, and $s_{+p}$ is the share of country $p$ in the sum of total export value (over all products and countries). Based on these RCA values, a product-by-country matrix **X** of dimensions $n \times m$ is constructed, in which element $x_{qp}$ is equal to 1 if $R_{qp} > 1$ and 0 otherwise. Thus, **X** is a binary RCA matrix that represents the information on trade performance of the countries.

Matrix **X** is a two-way table indicating which products are present/absent in the export specialization portfolio of which country. For every country (product), this table contains $n$ ($m$) dimensions of binary data. The essential information in this dataset (which can as well be seen as a matrix representation of a bipartite network graph) is the pairwise similarity or difference between the countries in terms of their respective export specialization portfolios, and also the pairwise similarity or difference between the products in terms of the countries that are specialized therein.

Economic complexity analysis (1) is an application of CA (11-12) that reduces the dimensionality the co-occurrence information in this two-way table to just one indicator for the products (product complexity, PCI), and one for the countries (the Economic Complexity Index, ECI). The original idea in (1) was to derive complexity and the ECI by an iterative procedure called the "method of reflections". This starts from the **X** matrix and, in the first step, both row- and column-sums are calculated. The sum of a column is the number of products for which the country has a comparative advantage, and this is called "diversity" of the country. The sum of the row is the number of countries that have comparative advantage in the product, and this is called "ubiquity" of the product.

Each iteration of the method of reflections produces a country indicator and a product indicator. The initial values are diversity and ubiquity, respectively. Then in each next iteration, the country indicator is updated to the average of the product indicator of the previous iteration, over all products for which the country has comparative advantage. Similarly, the product indicator is updated to the average of the country indicator of the previous iteration, over all countries that have comparative advantage in the product. Note that CA can also be operationalized by an iterative procedure similar (though not identical) to the method of reflections, and referred to as "reciprocal averaging" (16).

Later on, the method of reflections was reformulated as an eigenvalue problem, where ECI became the leading non-trivial eigenvector of a particular matrix that preserves the essence of the method of reflections, while the product complexity index, PCI, for any given product, reduces to the average ECI of the counties with revealed comparative advantage in the given product (2). This eigen-analysis reformulation, which was later shown to be identical to CA (11-12), starts with two normalized versions of the RCA matrix **X**: a ubiquity-normalized version $\mathbf{X}^u$ where each element of **X** is divided by ubiquity of the row, and a diversity-normalized version $\mathbf{X}^d$ where each element is divided by diversity of the column. Then an $n \times n$ matrix is created as $\mathbf{C}^p = \mathbf{X}^d \mathbf{X}^{u\prime}$, where the prime denotes a transposition, and the $p$ superscript indicates that we have a product-by-product matrix. This matrix carries information about the (pair-wise) similarity between all products in terms of the communality of the countries that are specialized in them. This proximity metric can be seen as the opposite counterpart of the dissimilarity metric known as $\chi^2$ distance, in the sense that the contribution of a country (where both products are located commonly) to the metric is penalized (divisively) by the diversity of the country, as well as the divisive penalization by the respective ubiquity values of the product pair. Thus, matrix $\mathbf{C}^p$ represents a product space similar to the one pioneered by (17), although the latter draws on a proximity metric based on conditional probabilities of co-location.

The first eigenvalue of matrix $\mathbf{C}^p$ is 1, and the corresponding eigenvector is constant. The second eigenvector is used as a measure for product complexity (PCI). The sum of the eigenvalues, excluding the trivial first one, is equal to the trace of matrix $\mathbf{C}^p$, which represents the total variance of the trade data. Thus, considering a larger number of eigenvectors will capture a larger share of the variance of the trade data as transformed into $\chi^2$ distances. Also, Euclidean distances in the eigen-space that covers all non-trivial eigenvectors correspond exactly to $\chi^2$ distances between products in the full data, and spaces that use less eigenvectors provide an approximation to these distances (the more eigenvectors are used, the better the approximation). In other words, eigen-analysis yields a product-space that is reduced in dimensionality, meaning that PCI is essentially a one-dimensional product space.

The Economic Complexity Index (ECI) of a country can either be calculated as the second eigenvector of the alternative country-by-country matrix $\mathbf{C}^c = \mathbf{X}^{d\prime} \mathbf{X}^u$, or as the average of the complexity of the products (computed as the leading non-trivial eigenvector of $\mathbf{C}^p$) in which the country has a comparative advantage. The two methods are equivalent up to a multiplicative factor of the ECI, and, assuming $n > m$, the $m$ eigenvalues of $\mathbf{C}^c$ are also identical to the leading $m$ the eigenvalues of $\mathbf{C}^p$.

The higher-order eigenvectors of $\mathbf{C}^c$ or $\mathbf{C}^p$ capture a lower share of the variance in the underlying trade data, but in the dimension reduction exercise that CA is, they contain useful information that can potentially be exploited (van Dam et al., 2021). By including all eigenvectors, the entire original space can be re-built, or, conversely, excluding subsequent eigenvectors with ever-larger corresponding eigenvalues yields the best reduced dimensionality given the number of remaining eigenvectors. The interpretation of what each of those remaining dimensions (eigenvectors) means (i.e., what exact information it carries) must be done on the basis of which countries or products "load high" (or low) on the eigenvector (i.e., which cells of the eigenvector have higher or lower values). Such an interpretation must make use of information about the countries or products. For example, we may observe that the leading eigenvector of $\mathbf{C}^c$ tends to load high for developed countries, and low for developing countries, or that the leading eigenvector for $\mathbf{C}^p$ loads high for high-tech products.

Hiding country names and product descriptions, i.e., without some kind of external information about which countries are developed or which products are high-tech, interpretation of the summary dimensions (the eigenvectors or ordinations) would be impossible. In such a "blind" case, one could resort to choosing the sign of the eigenvectors of $\mathbf{C}^p$ such that they would be negatively correlated to product ubiquity (or the signs of the eigenvectors of $\mathbf{C}^c$ such that a positive correlation to country diversity results). This would yield an interpretation of the eigenvectors as generalized measures of ubiquity or diversity, as suggested in (1). But this interpretation was already debunked by (18), who rigorously proved the orthogonality (i.e., linear independence) of ECI and counties' diversity, which also implies the orthogonality of PCI and products' ubiquity as a corollary. Orthogonality to diversity and ubiquity applies also to the higher-order eigenvectors other than ECI and PCI.

CCA addresses this interpretability challenge by the incorporation of additional observable and codified information that is used instead of the informal information embodied in country names and product descriptions. For example, if we wish to relate the ordinations to development, we can include per capita GDP as an external variable. As will be explained in detail below, CCA will incorporate the information that this new variable brings in constructing the ordinations (eigenvectors). Also, instead of interpreting ex-post (trying to interpret each eigenvector after it has been extracted), CCA includes the information ex-ante, i.e., maximizing correlation between the country eigenvector(s) and the external variable(s), yielding as many usefully interpretable ordinations as there are external variables.

**CCA AS A MORE INTERPRETABLE ALTERNATIVE TO ECONOMIC COMPLEXITY**

The canonical component of CCA essentially consists of what the pioneering CCA publication (19) calls "interpretation of the ordination axes with the help of external knowledge and data" (the ordination axes are the reduced dimensions, i.e., product complexity and ECI in our application). In this way, CCA directly includes the "external knowledge and data" in the dimension reduction procedure: "the ordination axes are chosen in the light of known environmental variables by imposing the extra restriction that the axes be linear combinations of environmental variables" (19, p. 1167).

In the ecological tradition of Ter Braak (19), CCA is typically used to interpret the relation between physical attributes of a site, and the relative occurrence frequency (or the presence/absence) of a number of species that are under investigation. The non-canonical form of CA uses the co-occurrence of species at individual sites to quantify similarities between species. The canonical form adds to this picture a relation between the species and characteristics of the site, such as average temperature or humidity. Thus, the environmental site variables can be seen as an exogenous determinant of the relative species frequencies, and CCA summarizes this relationship into a low dimensional space into which both the species and the sites can be visualized.

CCA will construct a set of country-level indicators which, in the Hidalgo and Hausmann approach, are similar to ECI, and correspond to what in the ecological tradition are site scores. It will also construct a set of product-level indicators, which are similar to the ecological species scores, and to Hidalgo and Hausmann's complexity. While in the ecological tradition the environmental variables can be seen as explanatory variables, our counterpart of these variables, such as GDP per capita or economic growth, bears no such causal interpretation. All that CCA will identify in our case, are associations between different types of outcome variables: comparative advantage in products, and development-related variables such as GDP per capita (and the medium-run growth rates thereof), greenhouse gas emissions, urbanization etc.

As prescribed by (19), the country level CCA indicators will be linear combinations of the country variables, plus a constant. The crucial idea of CCA is that we wish to maximize the correlation between these linear combinations and a country indicator where each country's observation is equal to the average of the product indicator over all products for which the country has comparative advantage. (19) achieved this by the incorporation of a weighted multivariate regression into the reciprocal averaging algorithm. An overview of this iterative algorithm can be found in the Methods and Data section below. However, it was also underlined by (19) that the iterative algorithm can be formulated in terms of eigen-analysis, just as the method of reflections was abandoned in favor of the equivalent eigenvalue problem.

In this paper, we also use an eigen-analysis formulation of CCA. This is explained in detail in the Methods and Data section along with a discussion on the equivalence to the iterative method. Our method is based on the eigen-analysis of a matrix $\mathbf{\Phi} = \mathbf{YTC}^c$, where $\mathbf{C}^c$ is the same matrix as used in the economic complexity calculations (as explained in the previous section). Matrix $\mathbf{Y}$ is the matrix of (weighted-standardized) environment variables, in our case a number of country variables. Matrix $\mathbf{T}$ is an instrumental matrix. This is explained in detail in the Methods and Data section below, here it suffices to observe that the pre-multiplication of $\mathbf{C}^c$ with $\mathbf{YT}$ introduces the diversity-weighted regressions on the country variables, which is the key difference between CA and CCA.

Given $z$ selected country variables, the eigen-analysis of matrix $\mathbf{\Phi}$ according to the procedure explained in the Methods and Data section, yields four matrices. The first one, $\mathbf{\lambda}$, is a diagonalized matrix of the leading $z$ non-trivial eigenvalues of $\mathbf{\Phi}$. The second one, $\mathbf{E}$ (which essentially stores the leading $z$ non-trivial eigenvectors of $\mathbf{\Phi}$) provides the "predicted" country scores in $z$ dimensions. Each column of $\mathbf{E}$ is a different linear combination of the $z$ country variables stored in $\mathbf{Y}$. The third one, $\mathbf{U}$ is the matrix of product scores. Each of the $z$ columns of $\mathbf{U}$ is the RCA-weighted averages of the elements of the corresponding column of $\mathbf{E}$ and analogous to Hidalgo and Hausmann's product complexity. The fourth one, $\mathbf{V}$, is the matrix of country scores, where each of the $z$ columns is the RCA-weighted averages of the elements of the corresponding column of $\mathbf{U}$, analogous to the ECI. Matrices $\mathbf{U}$ and $\mathbf{V}$, which form the basis of our analysis, are computed by the transformation of matrix $\mathbf{E}$.

It is important to observe that a major difference between CA and CCA is the presence of two alternative country ordinations (i.e., $\mathbf{E}$ and $\mathbf{V}$) in CCA. As explained in the Methods and Data section, this is an implication of the multi-variate regression in CCA. The predicted values from this regression are integrated into the reciprocal averaging process. One can use either one of these two country ordination in the analysis of the results in terms of special scatterplots known as biplots (13).

**BIPLOTS AND INERTIA**

The CCA method typically aims at depicting site (country) scores and species (product) scores together in 2-dimensional scatter plots (i.e., using two of the canonical axes that were constructed) along with indications of how the site/species ordinations relate to the 'environmental' (in our case, "country") variables. These scatterplots are referred to as "biplots". We also present results in terms of such biplots.

A biplot is essentially two scatterplots overlayed. The first one plots products in terms of two selected columns of $\mathbf{U}$ (i.e., the selected product ordinations) against each other. This can be seen as a 2-dimensional representation of the full product space (represented by $\chi^2$ distances), specific to the aspects of development that we chose to include by the country variables. The second one positions the countries in this product space, just as the RCA matrix $\mathbf{X}$ maps countries in the full product space. For plotting countries in terms of the selected ordinations, we use the corresponding columns of $\mathbf{V}$. Selecting the first and second ordinations for the biplot will maximize the share of the variance in the underlying trade data that is "captured" in the biplot. Biplots displaying other ordinations (i.e., the third column and beyond) may also prove to be useful. Thus, the biplot provide a product space of reduced-dimensionality, and enables us to position countries in this space, thus facilitating conclusions about the relation of both products and countries to the external development variables.

Under the topic of "scaling", (13) suggests alternative ways to overlay countries and products in a consistent way in the biplot. An important question is whether one plots country scores as the RCA-weighted average of product scores, or vice versa. The technical term here is the concept of a "barycenter" which specifies whether countries will show up in the biplot as the inner space of the products, or vice versa. Among the alternatives suggested (e.g., see 13 for a thorough discussion), the particular scaling scheme (i.e., the so-called Type1 scaling) that we adopt in the construction of biplots uses $\hat{\mathbf{U}} = \mathbf{U}\boldsymbol{\lambda}^{-0.5}$ as the product score matrix, along with the country scores matrix rescaled as $\hat{\mathbf{V}} = \mathbf{V}\boldsymbol{\lambda}^{-0.5}$.

As explained in the Methods and Data section, the multi-dimensional country scores in $\mathbf{V}$ are computed as the RCA-weighted averages of the respective product scores in $\mathbf{U}$, which already implies that countries occupy the barycenter. The rescaling of each dimension separately by the square of the corresponding eigenvalue is necessary in order to make sure that the Euclidian distances between the countries (as depicted on a biplot) are interpretable as the best of approximation to the $\chi^2$ distances between the countries in the original RCA matrix. Nevertheless, since both the country and the products scores are rescaled by the same factor (in each dimension), the countries still remain the barycenter of the products, i.e., a country score on a given column of $\hat{\mathbf{V}}$ is the RCA-weighted average of the product scores in the corresponding column of $\hat{\mathbf{U}}$.

The third element that goes into the biplots as an additional overlay, is a representation of country variables by colored rays from the origin. These lines make it possible to interpret the position of either products or countries (or groups of them) in the biplot, by projecting the products or countries on these lines. The coordinates to which the country variable lines point, are computed as the diversity-weighted correlation coefficients between the country scores (belonging to the selected CCA dimensions) and the country variable to which the ray refers to. This is called the intra-class correlation coefficient (19), and can be interpreted similarly to a "loading" in principal components analysis. Thus, the slope of those lines is indicative of the relative importance of the country variable in each of the two CCA dimensions depicted on the biplot. A relatively flat (steep) slope indicates that the variable is associated mainly to the first (second) CCA dimension as plotted horizontally (vertically).

As we show in the Methods and Data section below, and similar to CA (18; 11), diversity and ubiquity are orthogonal to the country and product ordinations in CCA, respectively. Hence, they provide a distinct source of information on trade patterns that is not included in the standard biplots that are used for CCA. Given that these concepts play such an important role in the interpretation of economic complexity, we include those variables in the biplots, by varying the bubble size of the product groups that we plot proportionally to ubiquity, and the bubble size of countries, proportionally to diversity.

CCA maintains the interpretation of the eigenvalues representing a share of the variance of the trade data. Each $i^{th}$ eigenvector (where $i \in [1, z]$) captures a share of variance of the trade data given by the metric *Inert$_i$* (so-called "inertia"), which is the $i^{th}$ non-trivial eigenvalue as divided by $tr(\mathbf{C}^p) - 1$. In the interpretation of the results (biplots), and also benchmarking against the single non-canonical CA ordination suggested by Hidalgo & Hausmann's economic complexity analysis, the relative values of *Inert$_i$* can be related to the product and country scores on dimension *i*. As we present the eigenvalues in decreasing order of their corresponding eigenvalue, higher-order canonical dimensions will capture ever smaller shares of the total variance of the trade data. Note also that all *z* canonical axes together represent a total share of this variance that is smaller than 1, i.e., there remains a part of the variance that is not captured by the canonical analysis. Finally, we observe that the first *p* canonical dimensions usually capture a smaller share of the variance in the trade data than the first *p* dimensions of non-canonical CA, because the correlations to the country variables in canonical analysis present a restriction on the product and country scores.

**RESULTS**

We use data on the US$ value of trade for 2002, 2010 and 2018, with each year having a different version of the Harmonised System (HS) classification scheme. We use the 5-digit level, which yields slightly more than 5,000 products in each year. We use 5 different country-level variables for 146 countries: the log of GDP per capita in constant international $, the average annual compound growth rate of GDP per capita (over 8 years with the year of the trade data in the center), the urbanization rate, territorial GHG emissions in CO2 equivalent tons per capita, and the Voice and Accountability variable from the World Governance Indicators. Data sources and other details are discussed in the Methods and Data section.

Figure 1 illustrates the elements of a standard biplot for the year 2018. Although we have slightly more than 5,000 products and 146 countries, we plot only one product (HS290410, a particular kind of hydrocarbons produced in the organic chemistry industry) and one country (Italy), for illustrative purposes. The summary indicators ("dimensions") with the two largest eigenvalues that result from the analysis are on the axes of the figure, and the percentages displayed in the labels are the shares of inertia that are associated with each dimension (the corresponding value for the Hidalgo and Hausmann ECI is 4.1%, as will be shown below).

The colored lines represent the country variables. These point from the origin to the coordinates that represent the so-called intraclass correlation coefficients. For example, the dark blue line representing GDP per capita points far into the horizontal dimension, which implies that the CCA-1 dimension is strongly correlated to this variable. On the other hand, it points to a low value on the vertical axis, which implies a low correlation between CCA-2 and GDP per capita. Longer lines represent country variables that are stronger correlated with (at least one of) the CCA dimensions. The projection of a country on one of the colored lines, as indicated by the dotted lines for Italy, relates the country's score on the associated variable with its trade specialization pattern. We see that Italy projects positively on the GDP per capita, Governance and Growth lines, implying that the country tends to be specialized in exporting products that are typically exported with above average intensity by countries that have relatively high values for those variables. The projection of a product on one of the country variable lines works in a similar way. For example, HS290410 tends to be exported with above average intensity by countries with relatively high values for GHG emissions and Urbanization but relatively low growth. In this case, the growth line has to be extended "backwards", as indicated by the dashed part, to be able to visualize the "negative" projection.

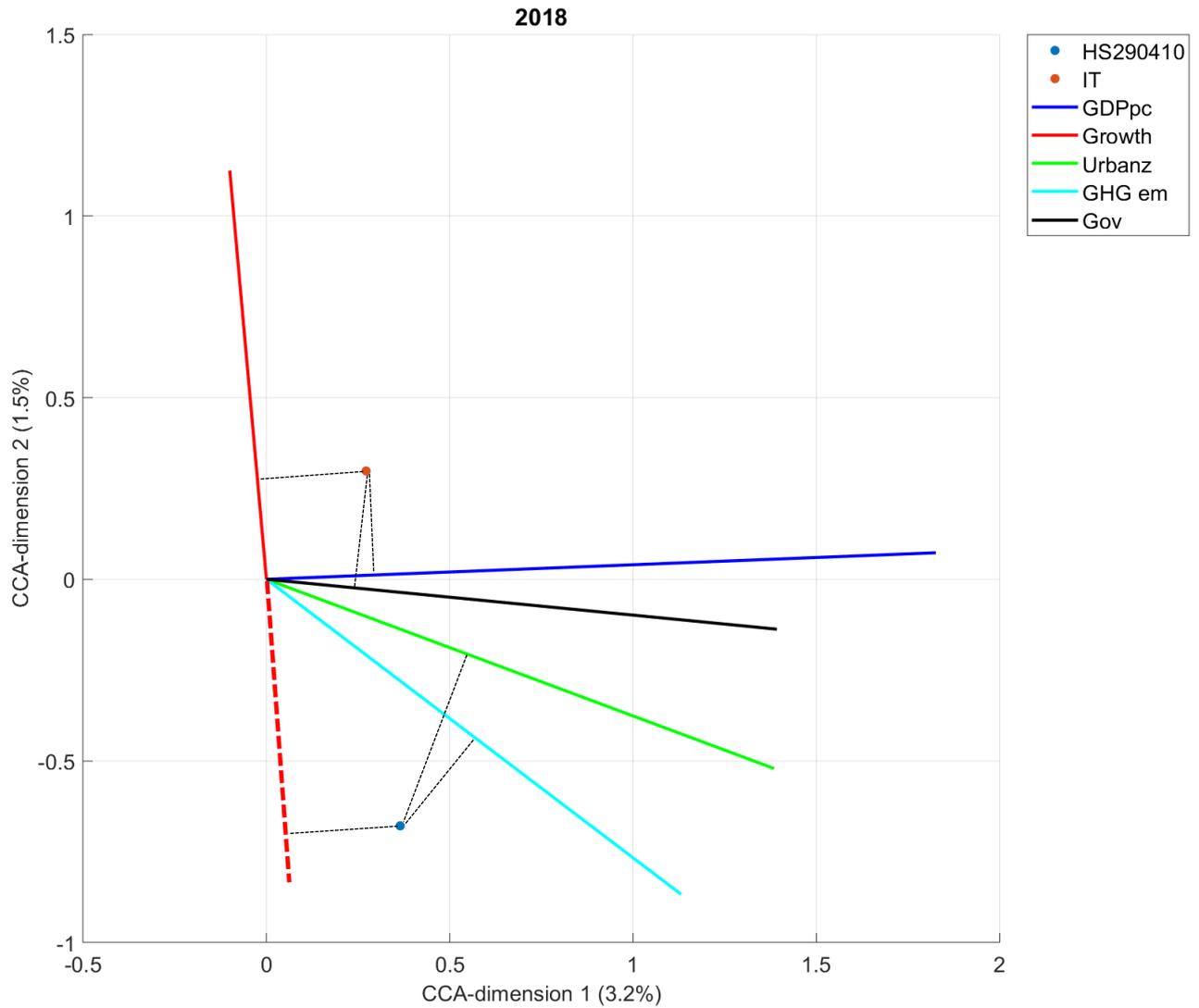

**Figure 1. Illustrative biplot**

In Figure 2, we compare CCA-1 with Hidalgo and Hausmann's ECI, for 3 years. As already announced, bubble size in these plots is varied proportionally to diversity of countries. In each of the years, ECI and CCA-1 are strongly correlated, even though ECI is associated to a slightly higher share of total inertia. Although diversity and the ordinations are orthogonal, there is a clear correlation between these ordinations and diversity: higher scores on either axis correspond to more diversified countries. The colored variable lines that were clearly distinguishable in Figure 1 now fold into a single dimension, where four of five variables are virtually on top of each other in terms of positive correlations with CCA-1 and ECI, and the (short) growth line points almost exactly in the opposite dimension, indicating a (weak) negative correlation. This "explains" why the literature that uses the ECI finds it easy to find a range of other variables that ECI is correlated to, although the relationship between Growth and ECI remains more problematic.

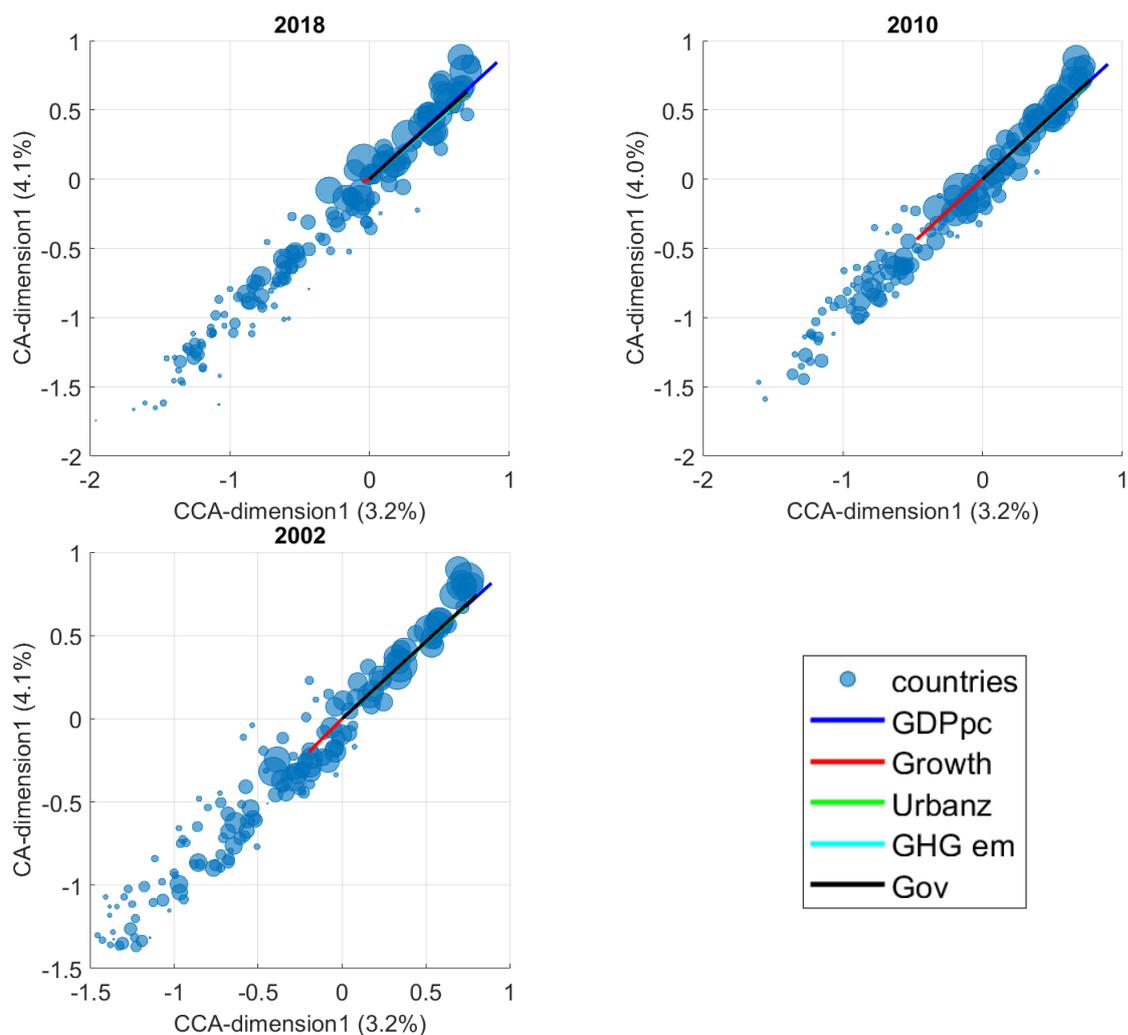

**Figure 2. CCA dimension 1 vs. ECI, 3 years**

Figure 3 shows the primary biplots, with the first two CCA dimensions, for all 3 years. Instead of the full set of 5,000+ products, we plot the centroids of 11 categories of products, which are listed in Table 1. This follows (20), who classifies products according to the level of technological capabilities that are necessary to produce and successfully export the products. Using the orthogonality between the product coordinates and ubiquity, we compute the Lall group centroids as the ubiquity-weighted average of the individual product coordinates in the biplot (the Methods and Data section provides details). Bubble size for the Lall product groups is varied proportionally to average ubiquity (for countries it is diversity as before). The lines corresponding to country variables are color-coded in a similar way as in Figures 1 and 2. Country labels are 3-digit ISO country codes, product labels are as in Table 1.

The correlation between the ubiquity of the product groups and CCA-1 is less clear than between countries' diversity and CCA-1 (Figure 2), but we do observe that the LTt, LTo, PPm, PPo and RBa groups have high ubiquity and are all on the lefthand side in the biplots (RBo is a clear exception to this pattern). Countries and Lall product groups that are positioned close to each other indicate countries' specialization patterns, and the projection of both product groups and countries on to the colored lines indicates the relationship between specialization patterns and the values of the country variables, as discussed above in relation to Figure 1.

| Lall category | Label |
|---|---|
| Primary products, minerals | PPm |
| Primary products, other (agriculture, forestry, fisheries) | PPo |
| Resource-based manufacturing, agriculture, forestry, fisheries | RBa |
| Resource-based manufacturing, other | RBo |
| Low-tech manufacturing, textiles etc. | LTt |
| Low-tech manufacturing, other | LTo |
| Medium-tech manufacturing, automotive | MTa |
| Medium-tech manufacturing, process-based (e.g., chemicals) | MTp |
| Medium-tech manufacturing, engineering-based (e.g., machinery) | MTe |
| High-tech manufacturing, electrical and electronics | HTe |
| High-tech manufacturing, other (e.g., pharma, aerospace) | HTo |

**Table 1. Lall product categories**

For example, in Figure 3A, we find Cambodia (KHM), Viet Nam (VNM) and Bangladesh (BGD) close to the LTt product group, and Saudi Arabia (SAU), Russia (RUS) and Kazakhstan (KAZ) close to the PPm group, and Zimbabwe (ZMB), Uganda (UGA) and Sierra Leone (SLE) close to the PPo group, which are all indications of the broad specialization patterns of these countries. We also find PPo, PPm, RBa, RBo and LTt in the periphery of these biplots, and the other groups (all medium- and high-tech, as well as the LTo) in the domain that projects positively on the Growth line as well as the 4 other variable lines. LTt also projects high on the growth axis.

By using CCA instead of CA, we obtained the colored variable lines, which will help in the interpretation of the ordinations. In Figure 3, the four variables other than Growth are always strongly and positively correlated to CCA-1, with much weaker correlations to CCA-2. Thus, as was already anticipated in the discussion of Figure 2, any interpretation that is facilitated by the four variables other than Growth, will have to come from higher-order ordinations, which in the case of Figure 3 is only CCA-2. The Growth line is always positively correlated to CCA-2, but negatively and usually (2018 and 2002) weakly correlated to CCA-1, which is consistent with the role of low-tech products and especially textiles in the process of "catching up" (e.g., 21).

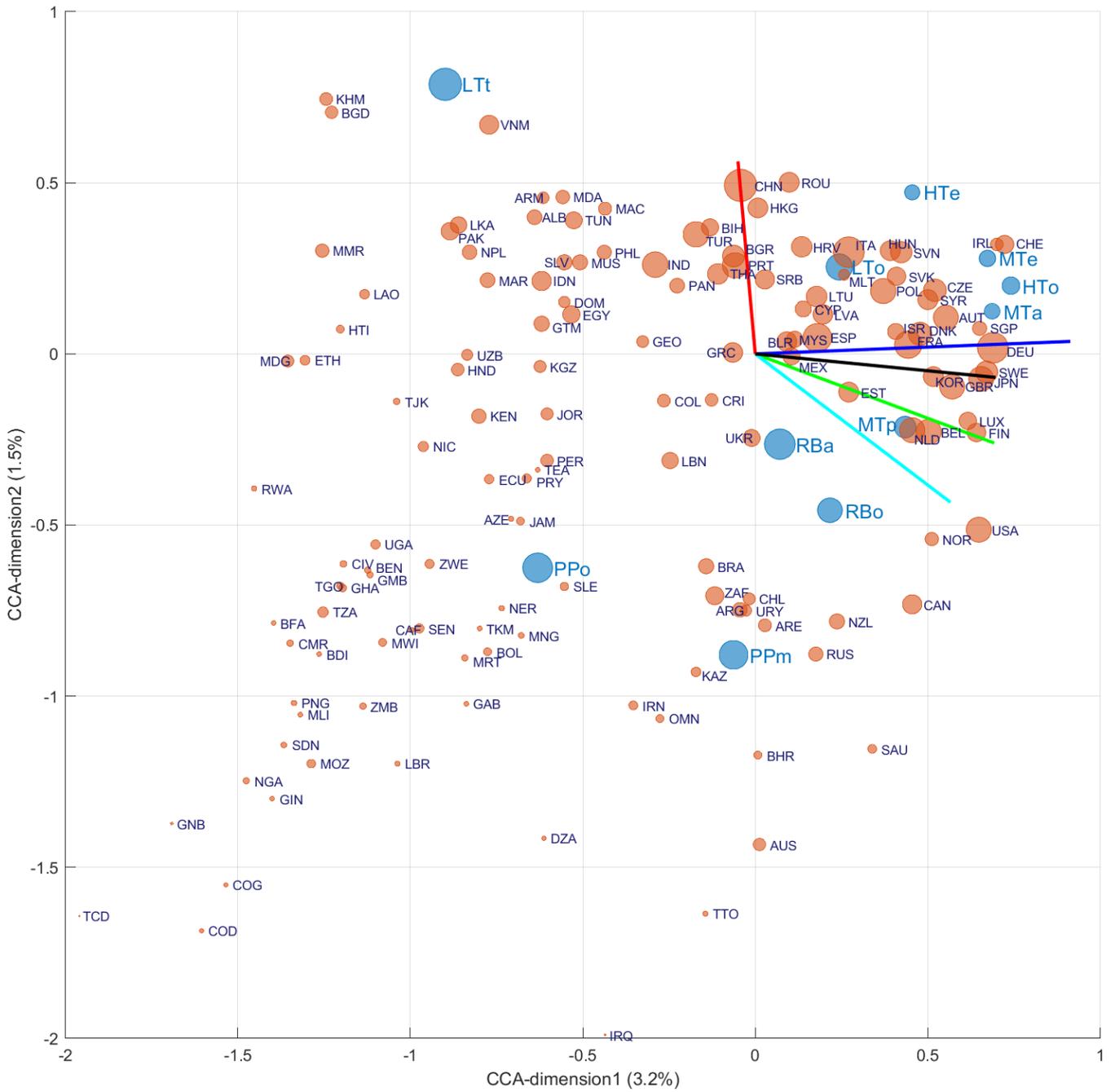

**Figure 3A. Biplots with Lall product groups, 1st and 2nd CCA dimensions, 2018**

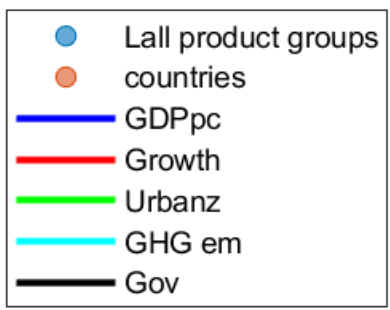

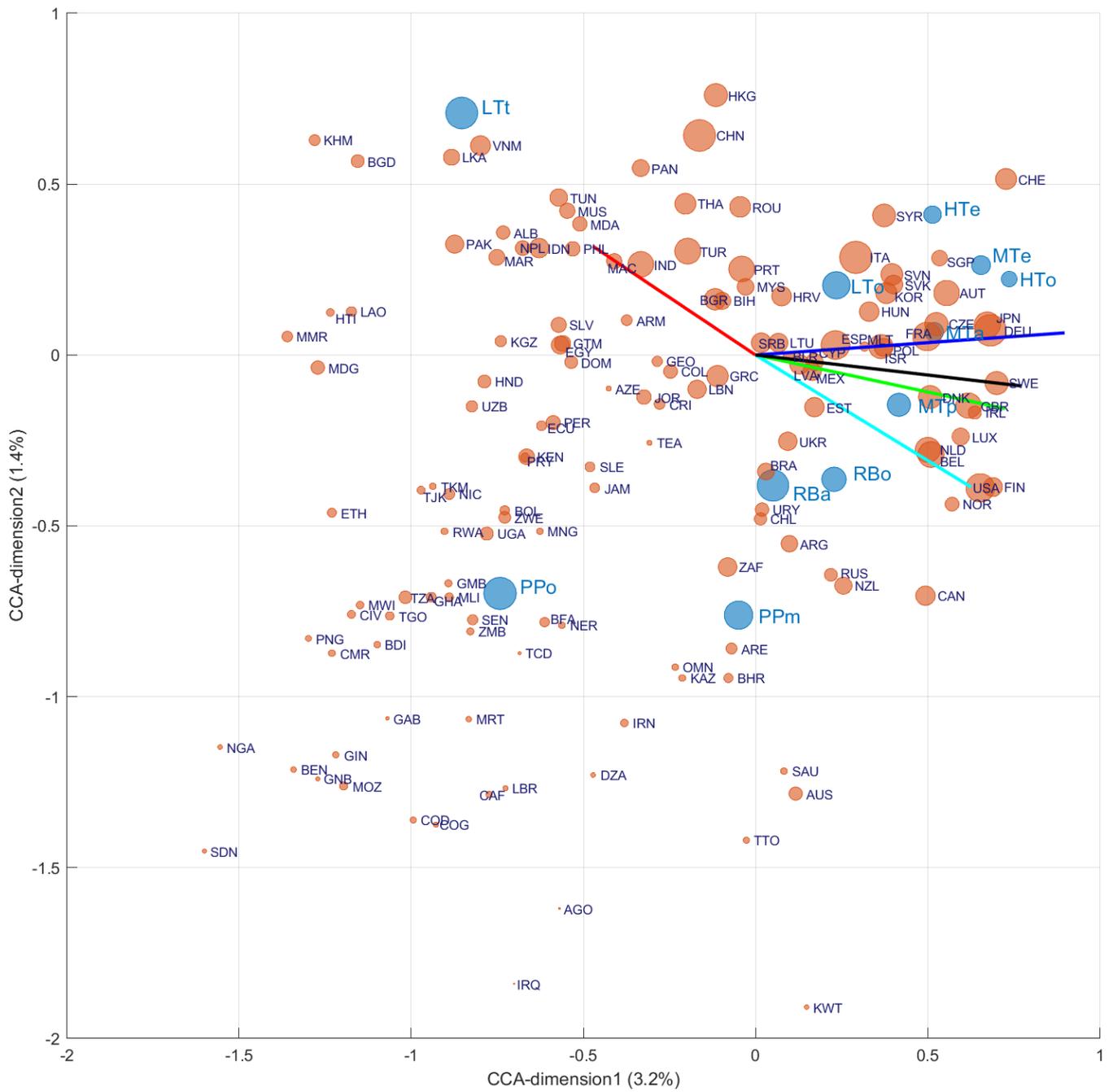

**Figure 3B. Biplots with Lall product groups, 1st and 2nd CCA dimensions, 2010**

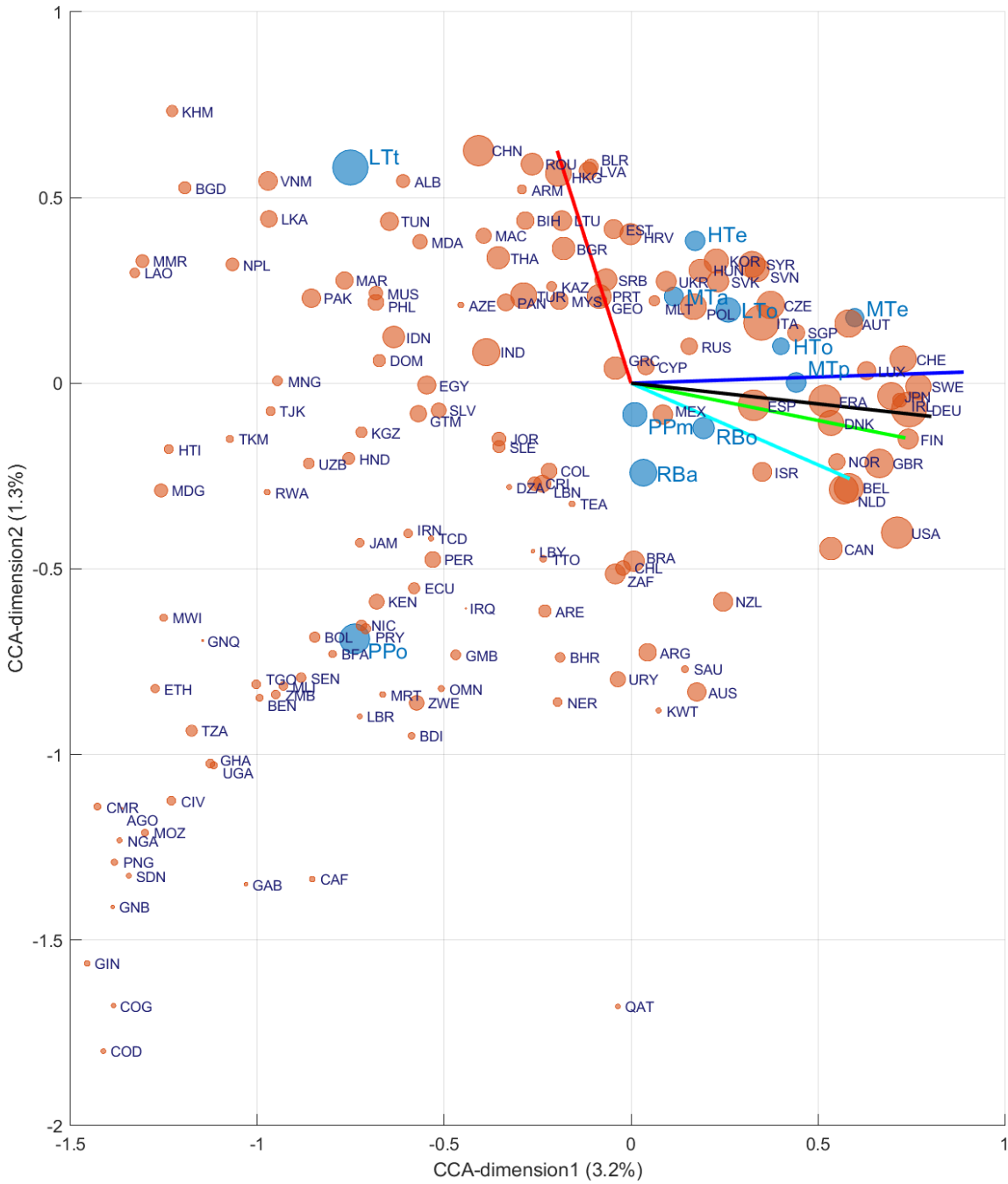

**Figure 3C. Biplots with Lall product groups, 1st and 2nd CCA dimensions, 2002**

In order to further explore the range of potential insights from CCA, we therefore construct biplots where CCA-1 is dropped and CCA-3 is introduced. Figure 4 shows these biplots. The CCA-1 dimension and its strong correlation to the country variables except Growth is now filtered out of the picture. The GDP per capita variable has the shortest variable line (while it was always the longest line in Figure 3). Therefore, we interpret the biplots in Figure 4 as indicative of a partial relationship between trade specialization patterns and country characteristics that abstracts from the correlation between the general development level of countries and trade specialization. Because the general development level of countries seems to be the strongest factor that identifies trade specialization (CCA-1 is constructed from the leading eigenvector), the biplots in Figure 4 capture a lower share of total inertia than those in Figure 3.

With GDP per capita (for the most part) filtered out, Governance, GHG emissions and Growth are now the variables with the strongest correlations, and they point into three separate directions, dividing the 360° circle roughly into 3 equal parts. This changes the general impression from the ordination considerably. For example, we now see a much more peripheral position for the PPm product group in the biplot, projecting very high on the GHG emissions variable line: when we abstract from the general correlation between development and GHG emissions, the latter becomes very strongly associated with the export of fossil fuels. In fact, we had to cap the upper axis limit for CCA-3 in Figure 4 to 1.5 to keep the rest of the figure legible, and this eliminates a number of countries that are all oil (and/or gas) exporters.

Another example of the partial nature of the biplots in Figure 4 is the fact that many (African) developing countries (e.g., Malawi, MWI; Sierra Leone, SLE; Senegal, SEN; Tanzania, TZA) project very high on the (black) Governance line. These countries do not stand out by their high values for the Governance variable, but once we eliminate the general correlation between GDP per capita and the Governance variable, we find that relative to their development level, these countries have high Governance values.

In terms of how the elimination of the CCA-1 axis affects the positions of individual countries in the biplot, Russia is a good example. In Figure 3C, Russia (RUS) projects close to the middle (although positively) on the Growth line, while in Figure 4C, it projects very high on Growth. This corresponds to the strong increase of the price of crude oil during the period 1998 – 2008, which is also reflected that in Figure 4C, the PPm product group projects positively on the Growth line, while in Figures 4A and 4B, it projects negatively). Due to the price of oil and the opening up of Russian trade, Russia is indeed a relatively fast grower in the period of Figure 4C. In the other periods (biplots 3A/4A and 3B/4B), oil price changes are less spectacular and the position of Russia is not so strongly affected by the elimination of CCA-1.

These are all examples of how Figures 3 and 4 construct fundamentally different interpretations (reduced-dimensional spaces) of the trade data. While Figure 3 captures more of the underlying variance (roughly 1.85 times more than Figure 4), both figures provide viable and interesting dimension reductions. The researcher may choose their own version, depending on the nature of the research questions.

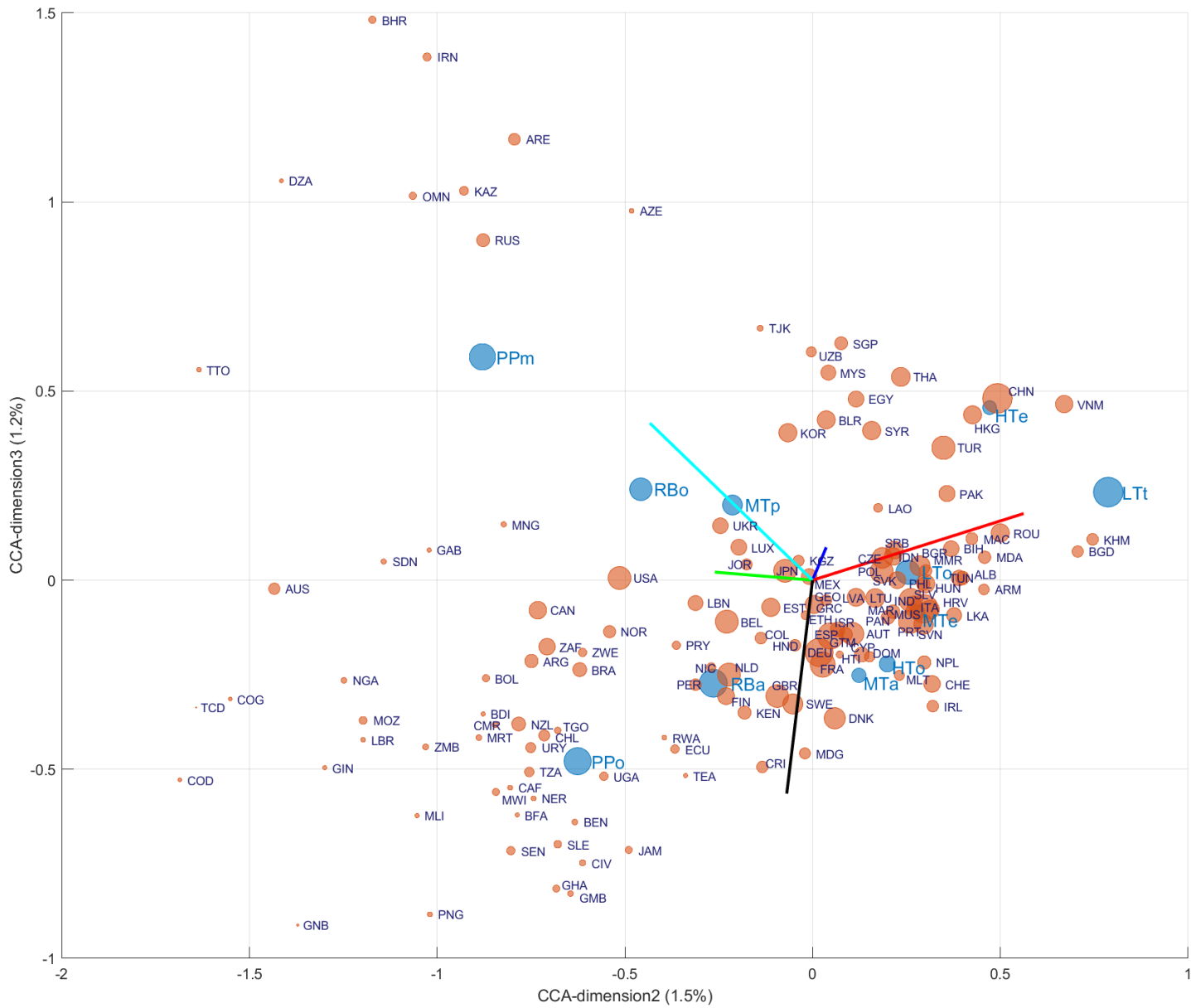

**Figure 4A. Biplots with Lall product groups, 2nd and 3rd CCA dimensions, 2018**

**Figure 4B. Biplots with Lall product groups, 2nd and 3rd CCA dimensions, 2010**

**Figure 4C. Biplots with Lall product groups, 2nd and 3rd CCA dimensions, 2002**

## DISCUSSION

Using CCA instead of plain CA to calculate a "complexity" index for countries' international trade specialization profiles has the main advantage that the index can be clearly interpreted in terms of macroeconomic variables that a researcher chooses to include. This also makes it easier to use multiple dimensions of the "complexity" indicator, because each dimension has its own interpretation in terms of the macroeconomic variables and can therefore be distinguished clearly from other dimensions. The single-dimension (CA-based) Economic Complexity Index (ECI) correlates strongly and positively to a range of indicators, as is shown in our analysis as well as in the literature at large. Using higher-order complexity indicators enables us to go beyond this general correlation between complexity and economic development, and be more specific about which (groups of) products are associated with which particular dimensions of development. The use of biplots, which are a graphical representation of the product space, is an important tool to help the interpretation of such multi-dimensional complexity indicators, also for the non-canonical version of CA.

Economic growth is an important variable in this context, as it is an indicator of the dynamics of development rather than the level of development. Growth is a volatile indicator, which can change rapidly between periods. Our results show that the correlation between Growth and ECI, or the first dimension from CCA, is generally much weaker than the correlations of development level variables to ECI. However, the second- and higher-order complexity indicators from CCA are much stronger correlated to economic growth. Thus, we can imagine to use the CCA-1 indicator as a way to "filter out" the level-development effects and use the higher-order CCA dimensions to further explore the Complexity – Growth relationship (cf. our Figure 4). In this way, low-tech textiles and high-tech electronics products emerge as the product groups that are consistently most strongly associated with high rates of economic growth.

## METHODS AND DATA

### CCA as an iterative algorithm

The first important fact to realize in the context of CCA (differently than other statistical methods of dimensionality reduction) is the absolute necessity to use weights in the standardization of the country variables and also in the regressions that are part of the algorithm. The algebra behind this necessity is beyond the scope of this paper and can be found in the quantitative ecology literature where CCA is an important element of the analytical toolbox (e.g., 13).

Accordingly, the procedure pioneered by Ter Braak in (19) starts by the standardization of the environment (in our case, country) variables in such a way that their weighted mean is 0 and their weighted standard deviation is 1, where the weight for each country is proportional to its diversity (i.e., the number of products where it has RCA>1). The weights should add up to 1.

The iterative algorithm commences by initializing the scores on the country indicator by arbitrary but distinct values. From there on:

- Step (1) of the procedure calculates product scores as the average of the country scores over all countries that have comparative advantage in the product.
- In step (2), new country scores are calculated as the average of product scores of the products for which the country has comparative advantage.
- In step (3), a multivariate linear regression is performed where the country scores are the dependent variable and the country variables (plus a constant) are the explanatory variables. This is a weighted regression where for each country, the weight is proportional to its diversity while all weights add up to 1.
- In step (4), the predicted values of this regression are adopted as the new country scores.
- Finally, step (5) standardizes the country scores such that their weighted average is 0 and their weighted variance is 1.

After step (5), one returns to step (1), unless changes in country scores were smaller than a threshold (i.e., convergence was obtained), in which case the procedure stops. The regression in step (3) and the use of the regression results in step (4) are what distinguishes this algorithm from the method of reflections. To calculate multiple canonical axes, one needs to repeat the procedure for each axis. For the second and higher-order canonical axis, an additional step needs to be inserted after step (5), aimed at making the fitted country scores orthogonal to the previous axes.

**CCA reformulated as an eigenvalue problem**

It is important to note that there is more than one way to formulate Ter Braak's iterative CCA algorithm (19) in terms of an eigenvalue problem. The "mainstream" version covered in (13) is hard to relate intuitively to Ter Braak's original iterative algorithm (thus to non-canonical CA, and thereby to Hidalgo and Hausmann's method of reflections), which is why we opt for an alternative that follows the footsteps of (19) more closely (see below on the equivalence).

The crucial part of the canonical element of CCA (vis-à-vis CA) is the weighted regression in step (3) of Ter Braak's iterative algorithm. To represent this in the eigenvalue analysis, we first define a matrix $\mathbf{Y}$ which stores the country variables in weighted-standardized form. $\mathbf{Y}$ has dimensions $m \times (z + 1)$ where $z$ is the number of country variables (in our case, $z = 5$). One column of $\mathbf{Y}$ that is to be used as a constant in a regression is populated with 1s, hence the number of columns of $\mathbf{Y}$ is $z + 1$.

We also need an $m \times m$ diagonal weight matrix $\mathbf{W}$ with diversity of the country (i.e., to be obtained as the respective column sum of $\mathbf{X}$) divided by the sum of diversity of all countries on the main diagonal, and zeros otherwise. Then matrix $\mathbf{T} = [\mathbf{Y}'\mathbf{W}\mathbf{Y}]^{-1}\mathbf{Y}'\mathbf{W}$, which enables the inclusion of the weighted regressions in the eigen-analysis (19).

This eigenvalue version of CCA extracts all canonical dimensions simultaneously. Thus, the regression of step (3) uses a set of country scores (coming from step 2) represented in a matrix $\mathbf{V}$ (instead of a vector of country scores in each instance of the iterative procedure). Given the definition of matrix $\mathbf{T}$, $\mathbf{B} = \mathbf{TV}$ will provide the regression coefficients, where $\mathbf{B}$ is a matrix of dimensions $(z + 1) \times z$. The $z$ columns of this matrix specify separate sets of regression coefficients, one for each canonical dimension to be extracted. In other words, each column of $\mathbf{B}$ and/or $\mathbf{V}$ corresponds to one instance of the iterative procedure.

The core of CCA is formed by the eigen-analysis of the matrix $\mathbf{\Phi} = \mathbf{YTC}^c$. Matrix $\mathbf{C}^c$ is the same as used in the complexity calculations of Hidalgo and Hausmann, and the pre-multiplication of $\mathbf{C}^c$ with $\mathbf{YT}$ introduces the weighted regressions on the country variables, which is exactly what turns ordinary CA into the canonical form. The first (trivial) eigenvalue of $\mathbf{\Phi}$ is 1, as it is in the method of reflections (i.e., non-canonical CA), and the corresponding eigenvector is ignored. The next $z$ (non-trivial) eigenvectors are extracted to form a matrix $\mathbf{E}$ in which each column contains one eigenvector. Thus matrix $\mathbf{E}$ has dimensions $m \times z$, i.e., with as many columns as there are non-trivial positive eigenvalues. Each column forms a canonical axis, or canonical dimension, for the countries. As was the case with the CA version of the method of reflections, complexity is now a multi-dimensional indicator, but the number of dimensions (i.e., $z$) is much smaller than the number of countries $n$ and/or the number of products $m$. Each column of **E** corresponds to the vector that is computed at the step (4) of Ter Braak's iterative algorithm. In other words, the matrix **E** contains the "predicted" country scores.

Note that the scale (norm) of the eigenvectors is arbitrary in principle, and different algorithms may deliver eigenvectors with different scales (many algorithms will settle for eigenvectors with norm equal to 1). Similar to step (5) of Ter Braak's algorithm, country diversity weights are used to standardize each column of the eigenvector matrix **E** such that its weighted mean is 0 and weighted variance is 1. This standardized eigenvector matrix is denoted as $\mathbf{E}^{Std}$. We then follow the iterative procedure, moving to step (1) to calculate product scores as RCA-weighted averages of predicted country scores: $\mathbf{U} = \mathbf{X}^u \mathbf{E}^{Std}$ where **U** has dimensions $n \times z$. Each column of **U** represents product scores on another canonical axis. The procedure finalizes with the computation of the country scores (in line with step (2) of Ter Braaks' algorithm) as the RCA-weighted averages of the product scores, i.e., $\mathbf{V} = \mathbf{X}^{d\prime}\mathbf{U} = \mathbf{X}^{d\prime}\mathbf{X}^u \mathbf{E}^{Std} = \mathbf{C}^c \mathbf{E}^{Std}$, which highlights the relation (and difference) between "predicted country scores" $\mathbf{E}^{Std}$ and the "country scores" **V**. We use the latter (as rescaled), along with **U** (also as rescaled similarly) in our biplots. As indicated above, the rescaling for the biplots is based on the leading $z$ non-trivial eigenvalues of **Φ** that are stored in the diagonalized matrix **λ**.

**Equivalence of the Eigenvalue Problem and Ter Braak's Iterative Algorithm**

Imagine that step (2) of the $p^{th}$ iteration of the iterative algorithm led to $\mathbf{V}(p) = \mathbf{X}^{d\prime}\mathbf{U}(p) = \mathbf{X}^{d\prime}\mathbf{X}^u \mathbf{E}^{Std}(p-1) = \mathbf{C}^c \mathbf{E}^{Std}(p-1)$. Note that we use $(p)$ to denote the state of a vector variable at the $p^{th}$ iteration. Following through, step (4) will yield $\mathbf{E}(p) = (\mathbf{YTC}^c)\mathbf{E}^{Std}(p-1)$ at the $p^{th}$ iteration. Also considering step (5) (i.e., the standardization of $\mathbf{E}(p)$ into $\mathbf{E}^{Std}(p)$, which at the point of convergence, reduces to a mere matter of rescaling per column), the convergence criterion of the algorithm requires that $\mathbf{E}^{Std}(p) = \mathbf{E}^{Std}(p-1)$. In matrix algebra, this condition can conveniently be written as $\mathbf{E}(p) = \mathbf{E}^{Std}(p-1)\boldsymbol{\lambda}$, where **λ** is a diagonal matrix. This immediately leads to $(\mathbf{YTC}^c)\mathbf{E}^{Std}(p-1) = \mathbf{E}^{Std}(p-1)\boldsymbol{\lambda}$ which is obviously an eigenvalue problem where **λ** is a diagonalized matrix of the leading $z$ non-trivial eigenvalues of the matrix $\boldsymbol{\Phi} = \mathbf{YTC}^c$.

**On the orthogonality between CCA ordinations and diversity/ubiquity**

First, let us write country diversity (i.e., the column sums of **X**) and products ubiquity (i.e., the row sums of **X**) formally. In order to do that define the $1 \times n$ column summation vector $\overline{\mathbf{1}}_n$ and the $m \times 1$ column summation vector $\overline{\mathbf{1}}_m$. Thus, the diversity vector is calculated as $\mathbf{d}' = \overline{\mathbf{1}}_n \mathbf{X}$ and the ubiquity vector as $\mathbf{s} = \mathbf{X}\overline{\mathbf{1}}_m$. Let us refer to the diagonalized versions of these two vectors respectively as **D** and **S**. Accordingly, the diversity-normalized RCA matrix can be written as $\mathbf{X}^d = \mathbf{X}\mathbf{D}^{-1}$ and the ubiquity normalized version as $\mathbf{X}^u = \mathbf{S}^{-1}\mathbf{X}$. Also note that the matrix which is used as weight in the multivariate regressions and also standardization of the eigenvectors is simply $\mathbf{W} = \frac{1}{x_+}\mathbf{D}$ where $x_+$ is the sum of all elements of **X**, thus total diversity and also total ubiquity

Now remember that we collect the first $z$ non-trivial eigenvectors of **Φ** in matrix **E** and further standardize into the matrix $\mathbf{E}^{Std}$ such that, for each column j, we make sure that $\mathbf{d}'\mathbf{E}_j^{Std} = \mathbf{0}$ and also $\mathbf{E}_j^{Std\prime}\mathbf{W}\mathbf{E}_j^{Std} = \frac{1}{x_+}\mathbf{E}_j^{Std\prime}\mathbf{D}\mathbf{E}_j^{Std} = \mathbf{1}$.

First, we demonstrate the orthogonality of each product ordination to ubiquity which, for each column j of $\mathbf{U}$ requires $\mathbf{s}'\mathbf{U}_j = \mathbf{0}$. Substituting $\mathbf{U}_j = \mathbf{X}^u \mathbf{E}_j^{Std}$ which further implies $\mathbf{U}_j = \mathbf{S}^{-1}\mathbf{X}\mathbf{E}_j^{Std}$, one can immediately observe that $\mathbf{s}'\mathbf{U}_j = \mathbf{s}'\mathbf{S}^{-1}\mathbf{X}\mathbf{E}_j^{Std} = \overline{\mathbf{1}}_n \mathbf{X} \mathbf{E}_j^{Std} = \mathbf{d}'\mathbf{E}_j^{Std}$, which is indeed equal to 0 due to the standardization procedure explained in the previous paragraph.

These properties are important to consider for the construction of the biplots, especially the aggregation of products into Lall categories. The orthogonality between the product ordinations and ubiquity (i.e., $\mathbf{s}'\mathbf{U}_j = \mathbf{0}$ for each ordination j) implies that the origin of the biplot exactly corresponds to the ubiquity-weighted centroid of all products depicted. In order to preserve this property, one should perform ubiquity-weighted averaging for the computation of the coordinates (i.e., centroids) of the product-aggregates.

It is straightforward to show similarly that the equation that determines the country scores at the next step of our procedure (i.e., $\mathbf{V} = \mathbf{X}^{d'}\mathbf{U}$) implies that each of the z country ordinations is orthogonal to diversity, i.e., for any $j^{th}$ column of $\mathbf{V}$, $\mathbf{d}'\mathbf{V}_j = \mathbf{0}$. This property, further implies that, in our biplots, the origin coincides with the diversity-weighted centroid of all country coordinates.

**On the intraclass correlation coefficients**

The intraclass correlation coefficients are used as the coordinates of the colored rays that represent the country variables on the biplots. They are weighted correlations, calculated on weighted-standardized variables, and hence are different from the usual (Pearson) correlation coefficients. For any selected dimension $j = 1,2,..z$, the computation requires the weighted standardization of the $j^{th}$ column of $\mathbf{V}$ (i.e., $\mathbf{V}_j$), such that the standardized version ($\widetilde{\mathbf{V}}_j$) satisfies $\overline{\mathbf{1}}'_m \mathbf{W} \widetilde{\mathbf{V}}_j = \mathbf{0}$ and $\widetilde{\mathbf{V}}'_j \mathbf{W} \widetilde{\mathbf{V}}_j = \mathbf{1}$. Because the columns of matrix $\mathbf{Y}$ that holds our county variables, are already similarly standardized (i.e., $\overline{\mathbf{1}}'_m \mathbf{W} \mathbf{Y}_j = \mathbf{0}$ and $\mathbf{Y}'_j \mathbf{W} \mathbf{Y}_j = \mathbf{1}$) , the intra-class correlation between the $i^{th}$ country variable and the $j^{th}$ country ordination is computed as $\mathbf{A}_{ij} = \mathbf{Y}'_i \mathbf{W} \widetilde{\mathbf{V}}_j$.

**Data sources**

Our data were downloaded from publicly available resources. Data on trade by products are from the UN Comtrade database accessed through the World Bank's WITS server (https://wits.worldbank.org/). We use data on "mirrored imports" in US$, i.e., total world imports by individual partner country). Trade data were downloaded continuously over the years 2018-2024. Data on GDP per capita were downloaded from the World Bank's World development Indicators (WDI) database in May 2024. We use GDP per capita in constant 2021 international $. Growth rates of GDP per capita were calculated from the GDP per capita data. Data on Urbanization are also from the WDI, and also downloaded in May 2024. This is urban population as % of total population in a country. Data on the Voice and Accountability governance indicator were taken from the World Governance Indicators database (22). Data on GHG emissions were taken from the European Union's EDGAR database (23).

**Acknowledgements**

We thank Maria Savona and other participants at the "Catching up and global value chains in times of transformation" Final CatChain Symposium, 19-20 February 2024 in Maastricht, at the GGDC conference in Groningen, 28-29 May 2024, at the International Schumpeter Conference in Gothenburg, Sweden, 9-11 June 2024, and at the Economic Fitness and Complexity Summer School, Maastricht, 8-12 July, for comments on preliminary versions. The views expressed and any remaining error remain solely our responsibility.

**Data and code availability**

The Matlab scripts used to perform the calculations and the data (collected from publicly available resources) are available from the authors on request.